# Electrically pumped AlGaN edge-emitting UV-B laser diodes grown by molecular beam epitaxy

Huabin Yu[†], Shubham Mondal[†], Rui Shen, Md Tanvir Hasan, David He, Jiangnan Liu, Samuel Yang, Minming He, Omar Alkhazragi, Danhao Wang, Mackillo Kira, Parag Deotare, Di Liang, Zetian Mi[*]

Department of Electrical Engineering and Computer Science, University of Michigan, Ann Arbor, MI 48109, USA

[*]E-mail: ztmi@umich.edu

[†]These authors contributed equally to this work

## Abstract

Mid and deep ultraviolet (UV) laser diodes remain among the least explored devices in semiconductor optoelectronics, despite their importance for spectroscopy, biochemical sensing, disinfection, and emerging quantum photonics. Here, we demonstrate an electrically pumped AlGaN-based laser diode operating in the UV-B band (280-315 nm). The device is grown by molecular beam epitaxy (MBE) on single-crystal AlN substrate and fabricated in a ridge-waveguide geometry. The laser diode operates at 298.5 nm and exhibits a relatively low threshold current density of 3.4 kA/cm$^2$. Clear nonlinear light-current characteristics and pronounced spectral narrowing with a full-width-at-half-maximum (FWHM) of 0.2 nm are measured above threshold.

Ultraviolet (UV) semiconductor laser diodes based on the AlGaN heterostructures are widely regarded as essential building blocks for compact UV photonic systems, enabling applications ranging from biochemical sensing and fluorescence spectroscopy to disinfection, high-resolution lithography, UV metrology, and emerging quantum photonics [1-3]. Over the past decades, substantial progress has been made for electrically pumped AlGaN-based laser diodes operating in the near-UV spectral range of 340-400 nm, where continuous-wave operation and relatively low threshold current densities have been demonstrated [4-6]. However, extending electrically pumped operation toward shorter wavelengths remains challenging. In particular, electrically pumped laser diodes operating in the UV-B (280–315 nm) and UV-C (200–280 nm) have seen only limited experimental demonstrations to date, owing to some fundamental material challenges [7-12]. Compared with UV-A (315-400 nm) laser diodes, both UV-B and UV-C devices require significantly higher aluminum (Al) compositions in the AlGaN heterostructures, which result in drastically reduced p-type doping efficiency, increased electrical resistivity, and stronger polarization-induced internal electric fields [13,14]. These factors severely limit carrier injection efficiency, suppress radiative recombination efficiency and reduce the maximum achievable optical gain under electrical injection. Moreover, in contrast to UV-C laser diodes, UV-B laser diodes often suffer from the lack of pseudomorphic growth on available substrates [15,16], arising from the larger lattice mismatch between AlGaN active region and single-crystal AlN substrate. The resulting dislocations, alloy disorder, and strain relaxation further introduce additional nonradiative recombination, optical scattering and internal loss, leading to increased lasing threshold and performance degradation. Despite these challenges, UV laser diodes emitting below 315 nm are attracting increasing interest because of their direct overlap with intrinsic molecular absorption bands and their potential for realizing compact, on-chip UV light sources for sensing, metrology, water/air purification, disinfection, and quantum and defense-relevant photonic systems [2,17].

Early efforts toward short-wavelength UV lasers were largely limited to optically pumped devices [18-21]. A major conceptual advance was introduced through the use of III-nitride nanostructures, which enable strain relaxation and reduced extended defect densities. Using self-assembled Al(Ga)N nanowire heterostructures, electrically pumped mid- and deep-UV laser diodes, including emission down to 239 nm, were reported beginning in 2015 [11,12,22]. More recently, significant progress has been achieved in planar edge-emitting UV-B and UV-C laser diodes through the use of high-quality single-crystal AlN substrates, which substantially reduce threading dislocation densities in the device active regions [7,8,10,23]. Notable advances include electrically pumped edge-emitting AlGaN laser diodes operating at 271, 275, and 298 nm in the mid- and deep-UV spectral regions, although threshold current densities remain high (often several to tens of kA $cm^{-2}$). Moreover, the demonstration of electrically pumped edge-emitting laser diodes operating in the UV-B remain very limited [10,24], with reported threshold current densities ~40–70 kA $cm^{-2}$. Thus far, such edge-emitting UV-B and UV-C laser diodes have been realized exclusively using metal–organic chemical vapor deposition (MOCVD). Molecular beam epitaxy (MBE), on the other hand, offers several unique advantages for short-wavelength UV laser diodes that remain comparatively underexplored [25,26]. These include ultra-low background impurity incorporation, precise control of aluminum composition and atomically sharp heterointerfaces, and the ability to implement advanced heterostructure designs such as polarization-engineered superlattices, tunnel junctions, monolayer-scale quantum wells, and excitonic engineering schemes which may significantly improve the performance of UV laser diodes.

In this work, we demonstrate an electrically pumped AlGaN-based UV-B edge-emitting laser diode operating at 298.5 nm. The devices were epitaxially grown on single-crystal AlN substrates by MBE and fabricated in a ridge-waveguide geometry incorporating distributed Bragg reflectors (DBRs). The electrical characteristics and electroluminescence (EL) spectra of the fabricated devices were systematically investigated. A relatively low threshold current density of 3.4 kA $cm^{-2}$ was measured, above which clear signatures of stimulated emission including nonlinear light-current characteristics and pronounced spectral narrowing of a full-width-at-half-maximum (FWHM) of 0.2 nm were observed. These studies establish MBE as a viable and compelling platform for achieving low threshold, electrically pumped mid and deep UV laser diodes.

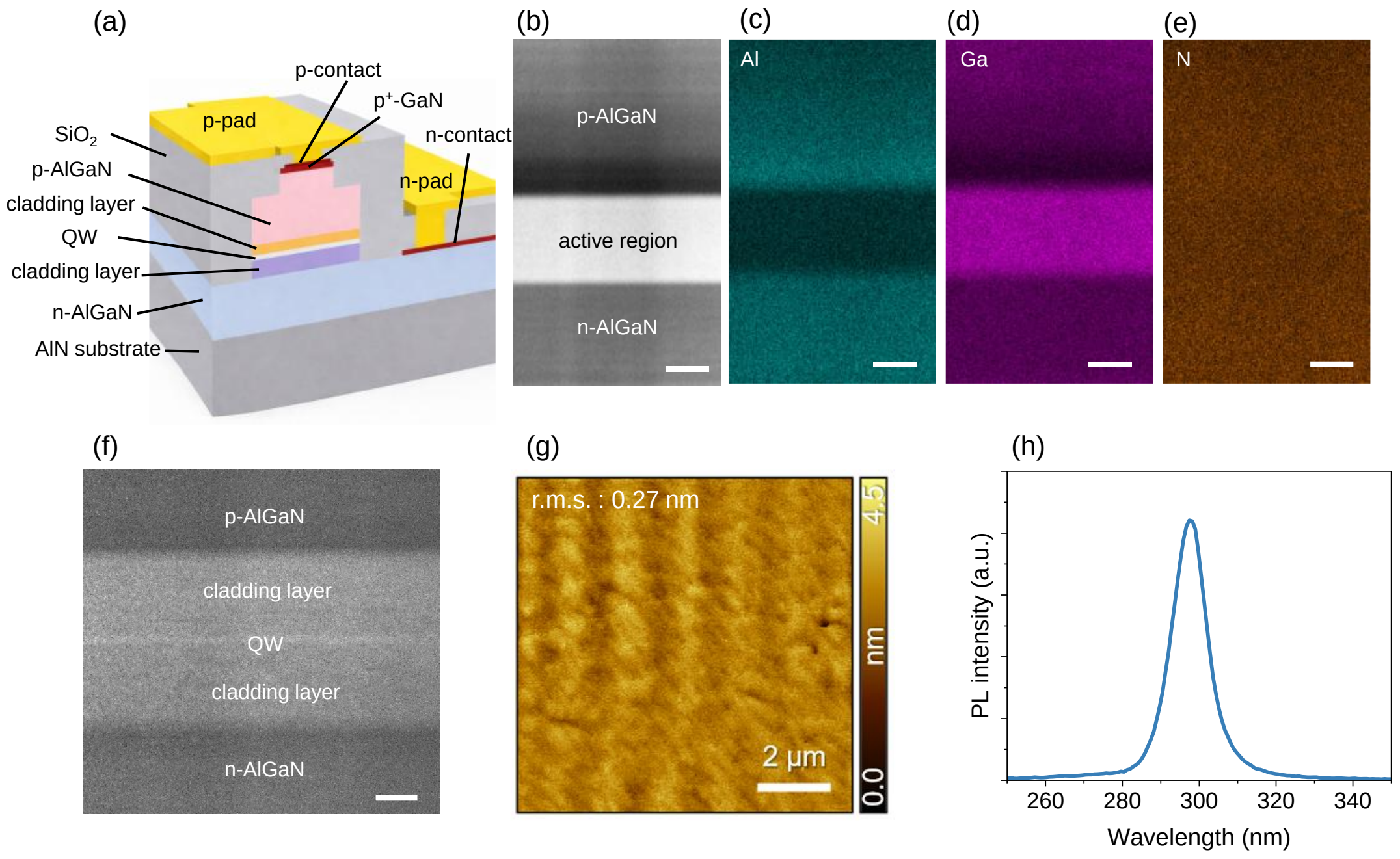

**Figure 1 Epitaxial structure and material characterization of the electrically pumped UV-B laser diodes.** (a) Schematic illustration of the electrically pumped AlGaN-based UV-B laser diodes. (b) Cross-sectional HAADF-STEM image of the laser heterostructure along the growth direction, clearly resolving the p-AlGaN cladding layer, active region, and n-AlGaN layer. Scale bar: 50 nm. (c–e) Corresponding EDS elemental mapping images of Al, Ga, and N, respectively. Scale bar: 50 nm. (f) High-resolution cross-sectional STEM image of the active region, showing a well-defined single quantum well embedded between the upper and lower cladding layers. Scale bar: 25 nm. (g) Atomic force microscopy (AFM) image of the surface, showing a roughness of 0.27 nm across a 10 μm × 10 μm scanning area. (h) Photoluminescence (PL) spectrum of the AlGaN-based UV-B laser diode measured at room-temperature.

The AlGaN-based laser diode heterostructures were grown on single-crystal AlN substrates by plasma-assisted MBE. **Figure 1** presents the epitaxial structure and material characterization of the electrically pumped UV-B laser diodes. **Figure 1a** schematically illustrates the AlGaN-based UV-B laser diode, which adopts a ridge-waveguide geometry with laterally separated p- and n-type contact pads. The laser heterostructure consists of a 100-nm-thick AlN buffer layer, followed by a 300-nm-thick Si-doped n-type $Al_{0.65}Ga_{0.35}N$ layer. A 50-nm-thick undoped $Al_{0.45}Ga_{0.55}N$ lower cladding layer was then grown, followed by a 3-nm-thick $Al_{0.35}Ga_{0.65}N$ single quantum well (QW). The active region is capped by a 50-nm-thick undoped $Al_{0.45}Ga_{0.55}N$ upper cladding layer. Above the active region, a 250-nm-thick Mg-doped p-type composition-graded $Al_{0.85\rightarrow0.45}Ga_{0.15\rightarrow0.55}N$ layer was grown to facilitate efficient hole injection. Finally, a 50-nm-thick heavily Mg-doped GaN layer was grown as the p-contact layer.

The structural quality of the laser active region was examined by cross-sectional scanning transmission electron microscopy (STEM). **Figure 1b** presents a high-angle annular dark-field (HAADF)-STEM image of the epitaxial stack along the growth direction, showing the p-AlGaN layer, active region, and n-AlGaN layer. To further verify the compositional uniformity of the heterostructure, energy-dispersive X-ray spectroscopy (EDS) elemental mapping was performed. **Figures 1c–e** show the corresponding Al, Ga, and N elemental mapping, confirming the intended AlGaN layer sequence and the absence of noticeable elemental segregation. Shown in **Fig. 1f**, a well-defined single QW heterostructure is clearly resolved, embedded between the upper and lower AlGaN cladding layers with atomically sharp and abrupt heterointerfaces. The surface morphology of the laser heterostructure was further studied using atomic force microscopy (AFM), as shown in **Fig. 1g**. The surface exhibits an atomically smooth morphology with a root-mean-square roughness of 0.27 nm over a $10 \times 10$ $\mu m^2$ scan area. This value is lower than the surface roughness typically reported for MOCVD- and MBE-grown AlGaN laser structures (∼0.7 nm) [27,28]. Such ultralow surface roughness is indicative of exceptionally high material quality and is critical for achieving low threshold operation. The optical quality of the active region was evaluated by photoluminescence (PL) measurements. **Figure 1h** shows the PL spectrum of the AlGaN-based structure measured at room-temperature, wherein a strong emission peak at ~297 nm is observed with a full-width-at-half-maximum (FWHM) of 10.7 nm, comparable to values reported for high-Al-content AlGaN quantum well structures emitting in the UV-B spectral range [29,30].

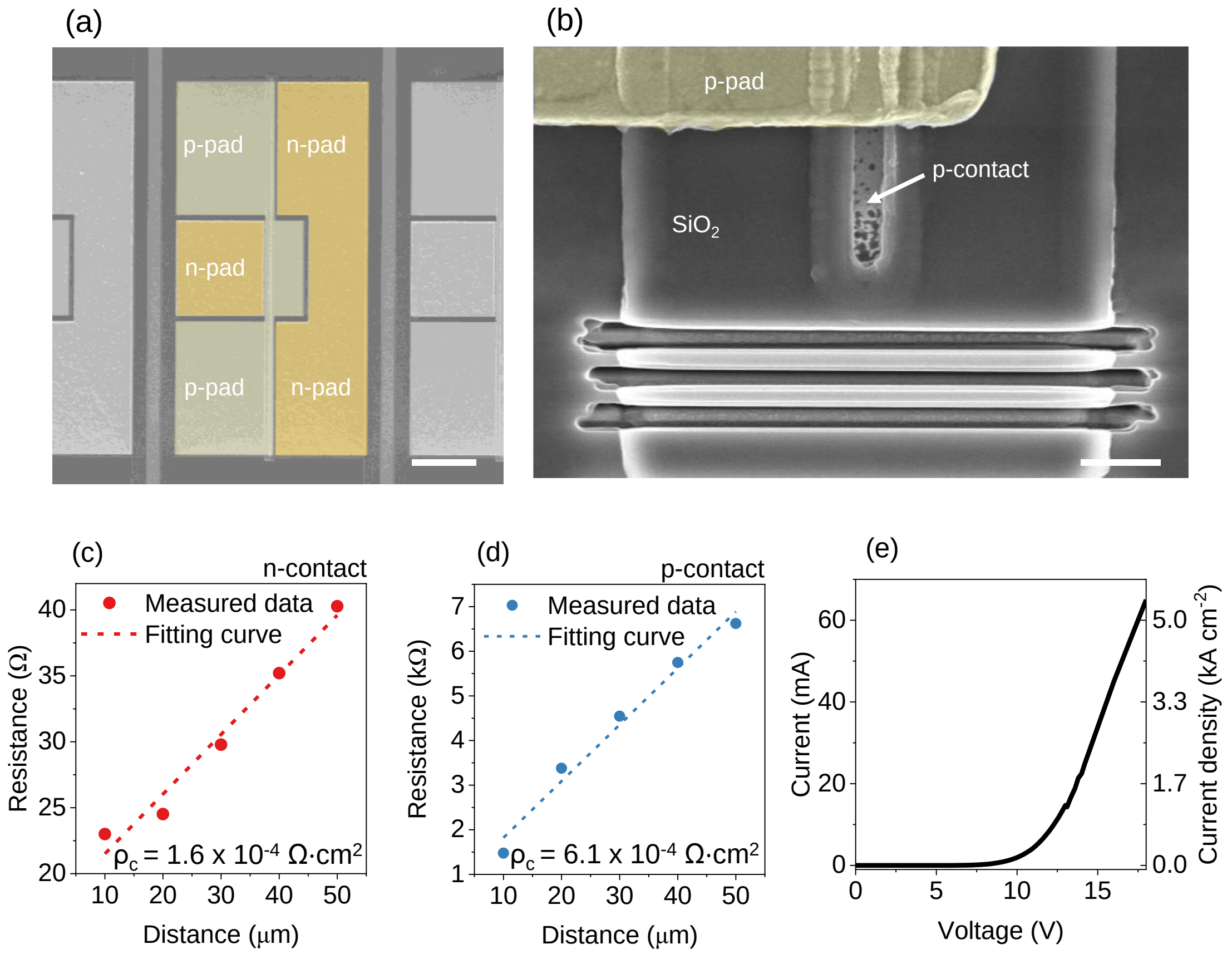

**Figure 2 Electrical characterization of the electrically pumped UV-B laser diodes.** (a) Top-view SEM image of the fabricated UV-B laser diode with defined p- and n-type contact pads. Scale bar: 100 μm. (b) SEM image of the focused-ion-beam (FIB)–defined DBRs patterned near the laser facet. Scale bar: 2 μm. Transmission line method (TLM) measurement of (c) n-type contact, and (d) p-type contact. (e) Current-voltage (I-V) and current density-voltage (J-V) characteristics of the device under electrical injection.

Electrically pumped UV-B laser diodes were subsequently fabricated utilizing standard photolithography, dry and wet etching, and contact metallization techniques. The laser heterostructure sample was first cleaned with ultrasonic treatment in acetone, isopropyl alcohol, and deionized water. Photolithography and dry plasma etching were used to define the ridge waveguide and device mesa, with etching performed down to the n-AlGaN layer for n-contact formation. Regions near the laser facets were further etched down to the AlN substrate to enable subsequent cavity engineering. The n-type (Ti/Al/Ti/Au) and p-type (Ni/Au) metal contacts were then deposited and annealed. To form the ridge waveguide, the p-type region surrounding the p-electrode was partially etched to approximately half of the p-AlGaN thickness. The p-electrode has a width of 2 μm, and a length of 600 μm. A 600-nm-thick $SiO_2$ passivation layer was subsequently deposited, followed by the deposition of a thick Ti/Au metal stack serving as bonding pads.

A top-view scanning electron microscopy (SEM) image of a representative fabricated device is shown in **Fig. 2a**, where the clear separation of the p- and n-contacts confirms the designed current-spreading scheme. Finally, focused-ion-beam (FIB) milling was employed to define distributed Bragg reflector (DBR) structures near the laser facet. Realizing high-quality laser facets in the deep- and mid-UV spectral range remains challenging due to the hardness of high-Al-content AlGaN, which limits the effectiveness of conventional dry etching techniques[16]. Moreover, the relatively low refractive index of AlGaN, compared to GaAs or InP based material family, further limits the maximum achievable reflectivity for as cleaved facets. To address these challenges, various approaches have been explored, including combined dry and wet etching of laser facets as well as cleaved facets integrated with atomic-layer-deposited (ALD) dielectric DBRs[8,23]. In the present work, we utilized FIB-defined DBRs to engineer and minimize facet-related optical losses in our UV-B laser diodes. **Figure 2b** shows a cross-sectional SEM image of DBRs defined by FIB etching. The thicknesses of the air gaps (~668 nm) and the AlGaN segments (~453 nm) are designed to satisfy the $(2k+1)\lambda/4$ quarter-wave Bragg condition at the target lasing wavelength, ensuring constructive interference of the reflected waves and enhanced optical feedback at the laser facet.

Electrical contacts to both p-type and n-type regions were characterized using the transmission line method (TLM). **Figures 2c and 2d** present the TLM results for the n-type and p-type contacts, respectively. The extracted specific contact resistivities are $1.6\times10^{-4}\ \Omega\cdot\mathrm{cm}^2$ for the n-type contact and $6.1\times10^{-4}\ \Omega\cdot\mathrm{cm}^2$ for the p-type contact, indicating relatively good ohmic contacts in both cases. These values are lower than some previously reported contact resistivities for high-Al-content AlGaN-based light-emitting devices, where achieving low-resistance ohmic contacts remains challenging due to limited dopant activation and wide bandgap barriers[31,32]. Further reduction of the overall series resistance can be achieved through optimized polarization-graded contact layers, refined metal stack engineering, improved annealing conditions, and the incorporation of tunnel junction.

**Figure 2e** shows the I-V and J-V characteristics under electrical injection. The device exhibits stable rectifying behavior with a monotonic increase in current density as the applied voltage increases. An injection current density of 3 kA/cm$^2$ can be achieved under a biasing voltage of around 15 V, which is still relatively high compared with long-wavelength III-nitride laser diodes but is comparable to previously reported electrically injected UV-B and UV-C AlGaN laser diodes [10,23], where large bandgaps, high series resistance, and limited dopant activation are common challenges. Further reduction of the operating voltage can be achieved in future studies through improved p-type doping incorporation, optimized polarization-graded injection layers, reduced contact resistance [2].

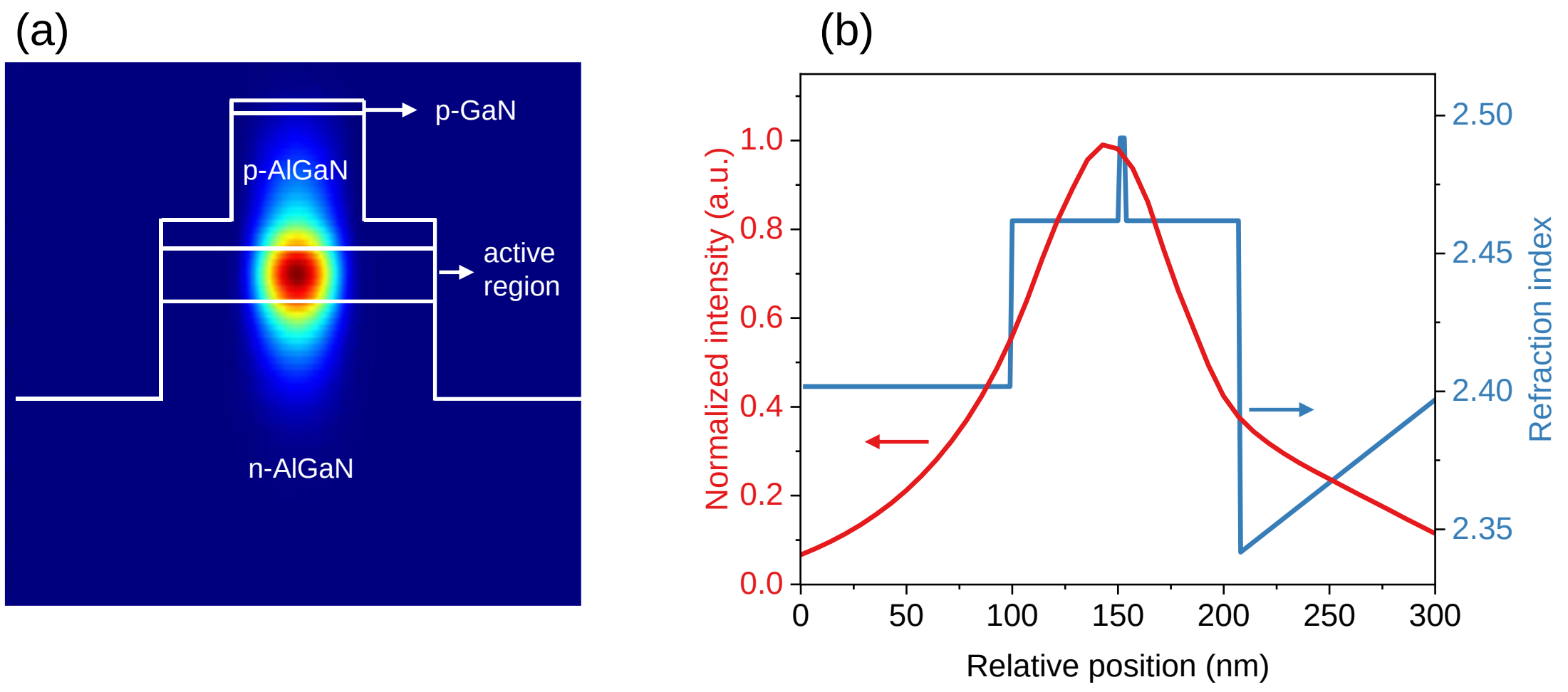

**Figure 3 Optical field distribution within the UV-B laser structure**. (a) Two-dimensional optical field intensity distribution calculated by finite-difference time-domain (FDTD) simulation, showing confinement of the optical mode within the active region of the ridge-waveguide laser diode. (b) Simulated normalized optical field intensity profile along the vertical direction of the device structure.

The optical mode characteristics of the UV laser diode were investigated by finite-difference time-domain (FDTD) simulation, as summarized in **Fig. 3**, with refractive indices and absorption coefficients of AlGaN and GaN taken from the literature and further confirmed by our own measurements [33,34]. Perfectly matched layers were applied to all simulation boundaries, and a single dipole source was placed at the center of the active region to excite the guided optical modes. **Figure 3a** shows the simulated two-dimensional optical field intensity distribution. The optical field is strongly localized within the active region, with the intensity maximum centered in the QW active region, indicating efficient overlap between the guided mode and the gain medium. Vertical confinement is provided by the AlGaN cladding layers, while lateral confinement is defined by the ridge-waveguide geometry, resulting in a well-confined optical mode with minimal leakage into the surrounding layers. **Figure 3b** presents the normalized optical field intensity profile along the vertical direction, further confirming that the modal energy is predominantly concentrated within the active region. From the field distribution, the optical confinement factor [35] of the QW is estimated to be approximately 1.4%. The simulated mode profiles indicate that the present device design provides sufficient optical confinement to support electrically pumped stimulated emission.

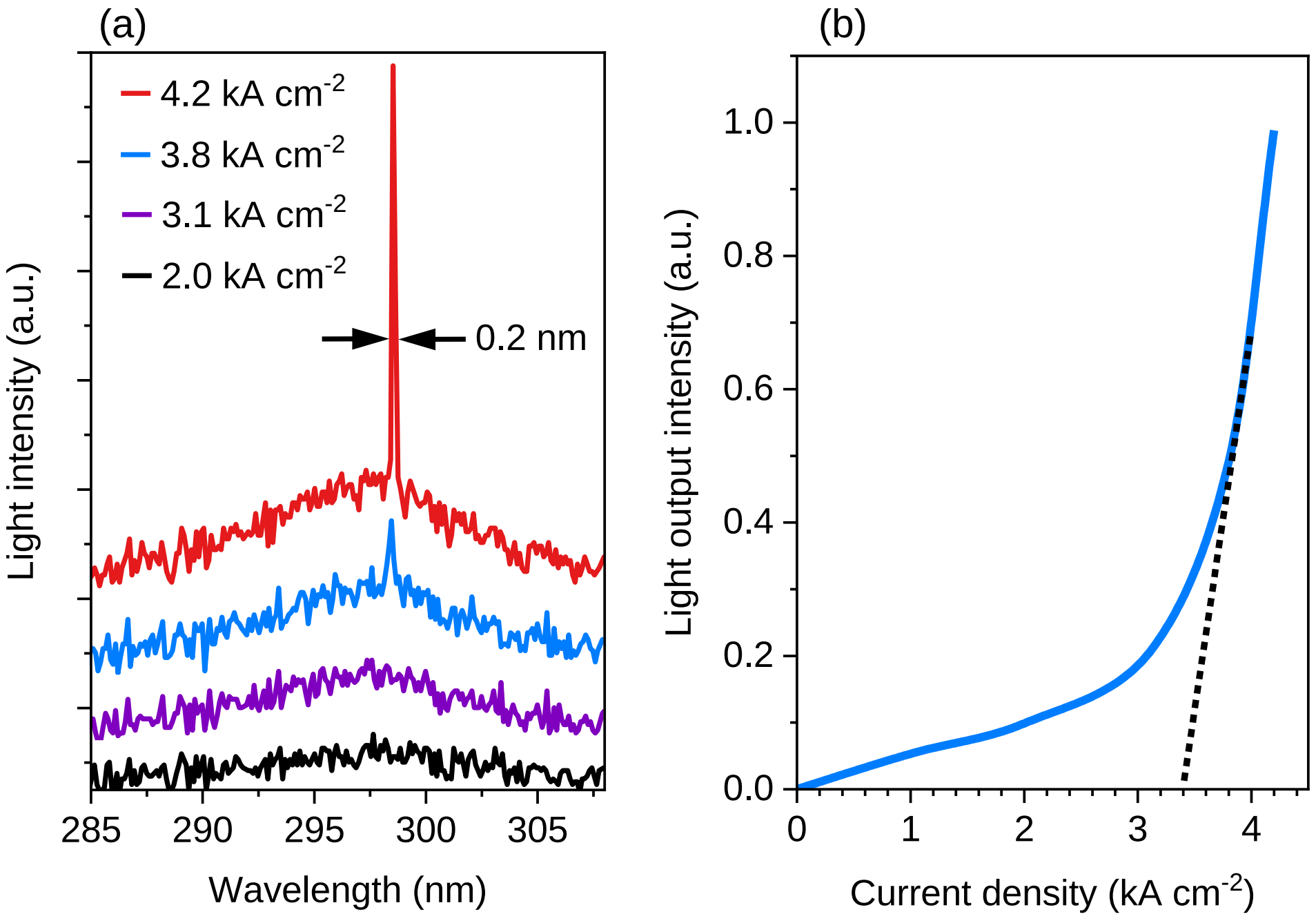

**Figure 4 Optical performance of the UV-B laser diode**. (a) Electroluminescence (EL) spectra recorded at different injection current densities. (b) Light–current (L-I) characteristics showing a clear threshold current density of 3.4 kA/cm$^2$.

The electrical and optical characteristics of the AlGaN-based ultraviolet laser diode are further investigated. All measurements were performed under pulsed operation (pulse width of 1 μs and duty cycle of 1%) at a stabilized heat-sink temperature of -15 °C. **Figure 4a** shows the EL spectra recorded at different injection current densities. The spectra were collected using an optical fiber and measured with a fiber-coupled spectrometer (Ocean Optics). At low current densities, the emission spectra are relatively broad and characteristic of spontaneous emission. A sharp emission peak emerges at approximately 298.5 nm can be clearly observed for the spectrum measured at an injection current of 3.8 kA/cm$^2$. With further increasing injection current, the emission peak remains stable at 298.5 nm and the intensity increases drastically. The measured spectral linewidth (FWHM) of the lasing peak is approximately 0.2 nm, which falls within the range typically reported for electrically pumped AlGaN-based UV-B laser diodes [10,24]. The L–I characteristics are presented in **Fig. 4b**. At low injection current densities, the emitted optical intensity increases gradually with current, consistent with spontaneous emission. As the injection current density increases near the laser threshold, a pronounced nonlinear increase in output intensity is observed, indicating the onset of stimulated emission. From the L–I curve, a threshold current density of approximately 3.4 kA $cm^{-2}$ is derived. This threshold current density is more than an order of magnitude lower than those of previously reported electrically pumped UV-B AlGaN laser diodes, which typically exhibit threshold current densities in the range of ~40–70 kA $cm^{-2}$ at similar lasing wavelengths [10,16]. Further reduction of the threshold current density can be achieved in future studies through suppression of nonradiative recombination, reduction of internal optical loss, improved current confinement, and the incorporation of advanced cavity engineering strategies such as optimized DBRs or index-guided ridge geometries [2].

In summary, we have demonstrated electrically pumped AlGaN-based UV-B laser diodes emitting at 298.5 nm. Detailed structural and materials characterization confirms the high crystalline and compositional quality of the epitaxial heterostructure. Clear nonlinear light-current characteristics, pronounced spectral narrowing with a linewidth of approximately 0.2 nm, and a threshold current density of ~3.4 kA $cm^{-2}$ were observed. Despite these advances, several challenges remain for further performance improvement, including the relatively high operating voltage, large series resistance associated with charge carrier transport and contacts, and optical losses related to cavity and facet engineering. Future efforts include improving p-type doping efficiency, reducing contact resistance, optimizing optical confinement and feedback, and incorporating advanced carrier injection schemes, which are expected to further reduce the threshold current density and operating voltage, ultimately enabling continuous-wave, high power operation in mid- and deep-UV laser diodes.

**Acknowledgements**

This work is supported by the Army Research Office under Grant No. W911NF2310142. The authors acknowledge the Lurie Nanofabrication Facility (LNF) and Michigan Center for Materials Characterization [$(MC)^2$] for use of the instruments and staff assistance. The authors also wish to thank Prof. Yiyang Li at the University of Michigan for help with some electrical characterization.

**Conflict of Interest**

Some IP related to this work has been licensed to NS Nanotech, Inc., which was co-founded by Z.

Mi. The University of Michigan and Mi have a financial interest in the company.